\documentclass{iau} 
\usepackage{graphicx}

\title[3-D visualisation of H\,{\small I} data ] 
{3-D interactive visualisation tools for H\,{\small I} spectral line imaging}

\author[J.M. van der Hulst, D. Punzo \& J. B. T. M. Roerdink]   
{J.M. van der Hulst, D. Punzo$^1$
 \and J. B. T.M. Roerdink$^2$}

\affiliation{$^1$Kapteyn Astronomical Institute, University of Groningen, \\ Landleven 12,
NL-9747AD, Groningen, the Netherlands \\ email: {\tt j.m.van.der.hulst@rug.nl \\
punzo@astro.rug.nl} \\[\affilskip]
$^2$Johann Bernoulli Institute for Mathematics and Computer Science, University of Groningen, \\ Nijenborgh 9, NL-9747AG Groningen, the Netherlands \\email: {\tt j.b.t.m.roerdink@rug.nl}}

\pubyear{2017}
\volume{325}  
\jname{Astroinformatics}
\editors{M. Brescia, G. Djorgovski, E. Feigelson, G. Longo \& S. Cavuoti, eds.}
\begin{document}

\maketitle
\begin{abstract}
Upcoming H\,{\small I} surveys will deliver such large datasets that automated processing using the full 3-D information to find and characterize H\,{\small I} objects is unavoidable. Full 3-D visualization is an essential tool for enabling qualitative and quantitative inspection and analysis of the 3-D data, which is often complex in nature. Here we present $\tt{SlicerAstro}$, an open-source extension of $\tt{3DSlicer}$,  
a multi-platform open source software package for visualization and medical image processing, which we developed for the inspection and analysis of H\,{\small I} spectral line data. We describe its initial capabilities, including 3-D filtering, 3-D selection and comparative modelling. 
\end{abstract}

\section{Introduction}

Upcoming H\,{\small I} surveys with instruments such as ASKAP (Australian Square Kilometre Array Pathfinder, \cite[Johnston et al. 2008]{2008ExA....22..151J}) and Apertif on the WSRT (Westerbork Synthesis Radio Telescope, \cite[Verheijen et al. 2009]{2009pra..confE..10V}, \cite[Oosterloo et al. 2010]{2010iska.meetE..43O}) will provide large data flows, in the regime of a few TB per day. Automation of the calibrating and imaging is imperative, and so is the automation of finding objects in the noisy H\,{\small I} data cubes. The expectation is that an H\,{\small I} survey with each of these instruments is capable of delivering of the order of 100 objects per day (\cite[Duffy et al. 2012]{2012MNRAS.426.3385D}). Automated source finders such as $\tt{Duchamp}$ (\cite[Whiting 2012]{2012MNRAS.421.3242W}) or $\tt{SoFiA}$ (\cite[Serra et al. 2015]{2015MNRAS.448.1922S}) will locate the detected objects. Yet in many cases, in particular when the objects are well resolved, further inspection is required to fully explore the characteristics of these object and interpret the subtle signs in the H\,{\small I} data of the various mechanisms of gas accretion and removal in galaxies (for an extended review of these see \cite[Sancisi et al. 2008]{2008A&ARv..15..189S}).

In the early 90's it was realised that viewers for radio-astronomical spectral line data would benefit greatly from advanced 3-D visualization software
\cite[(Norris 1994)]{1994ASPC...61...51N}.
It was very clear already then that full a 3-D approach would greatly help  understanding
of the 3-D structure of the radio data. The type of 2-D data slicing commonly used (i.e., channel movies), and implemented in packages such as $\tt{karma}$ \cite[(Gooch 1996)]{1996ASPC..101...80G}, is useful but limited in that it misses the capability to get a full 3-D view of the data. Recently, both \cite[Hassan \& Fluke (2011)]{2011PASA...28..150H} and \cite[Koribalski (2012)]{2012PASA...29..359K} pointed this out and and described need for 3-D visualisation capabilities in the context of upcoming H\,{\small I} surveys.

\cite[Punzo et al. (2015)]{2015A&C....12...86P} discussed the requirements for a fully interactive visualization tool: coupled 1-D/2-D/3-D visualization, quantitative and comparative capabilities, combined with supervised semi-automated analysis. In addition the source code must be open, modular, well documented, and well maintained. \cite[Punzo et al. (2015)]{2015A&C....12...86P} reviewed several existing packages and decided to develop a module for $\tt{3DSlicer}$ (URL: https://www.slicer.org/) which adds the following characteristics to the already large number of capabilities of the current $\tt{3DSlicer}$ system:  (i) the ability to read FITS (Flexible Image Transport System, \cite[Wells et al. 1981]{1981A&AS...44..363W}) files and handle the astronomical world coordinates system (WCS, \cite[Calabretta \& Greisen 2002]{2002A&A...395.1077C}, \cite[Greisen \& Calabretta 2002]{2002A&A...395.1061G}, and \cite[Greisen et al. 2006]{2006A&A...446..747G}), (ii) coupled 1-D, 2-D and 3-D display of the data and fast rendering, (iii) semi-automated filtering methods to recover faint emission using the information in all three dimensions (\cite[Punzo et al. 2016]{2016A&C....17..163P}), (iv) 3-D selection and (v) use of 3-D modelling to characterise and select the regularly rotating H\,{\small I} disks in galaxies. This paper gives a brief overview of the capabilities of $\tt{SlicerAstro}$. For a detailed description see \cite[Punzo et al. (2015)]{2015A&C....12...86P}, \cite[Punzo et al. (2016)]{2016A&C....17..163P} and Punzo et al. (2017, in preparation).

\section{$\tt{SlicerAstro}$}

$\tt{SlicerAstro}$ (URL: https://github.com/Punzo/SlicerAstro) is an open-source extension of $\tt{3DSlicer}$, provided via its built-in extension manager. All functionality within $\tt{SlicerAstro}$ is geared toward handling 3-D radio astronomical data, but makes use of all built in features of the core software of $\tt{3DSlicer}$, fully exploiting the fast 3-D rendering, linked display capabilities and segmentation or region of interest selection. It will be able to handle 3-D data cubes of op to $10^9$ voxels on modern desktop machines which are usually with rather fast GPU's.  The present implementation of $\tt{SlicerAstro}$ provides the following features that are of great interest for the exploration of radio astronomical spectral line (in particular  H\,{\small I}) data:
\newline
\begin{figure}[ht]
\vspace*{0.3 cm}
\begin{center}
 \includegraphics[width=5.2in]{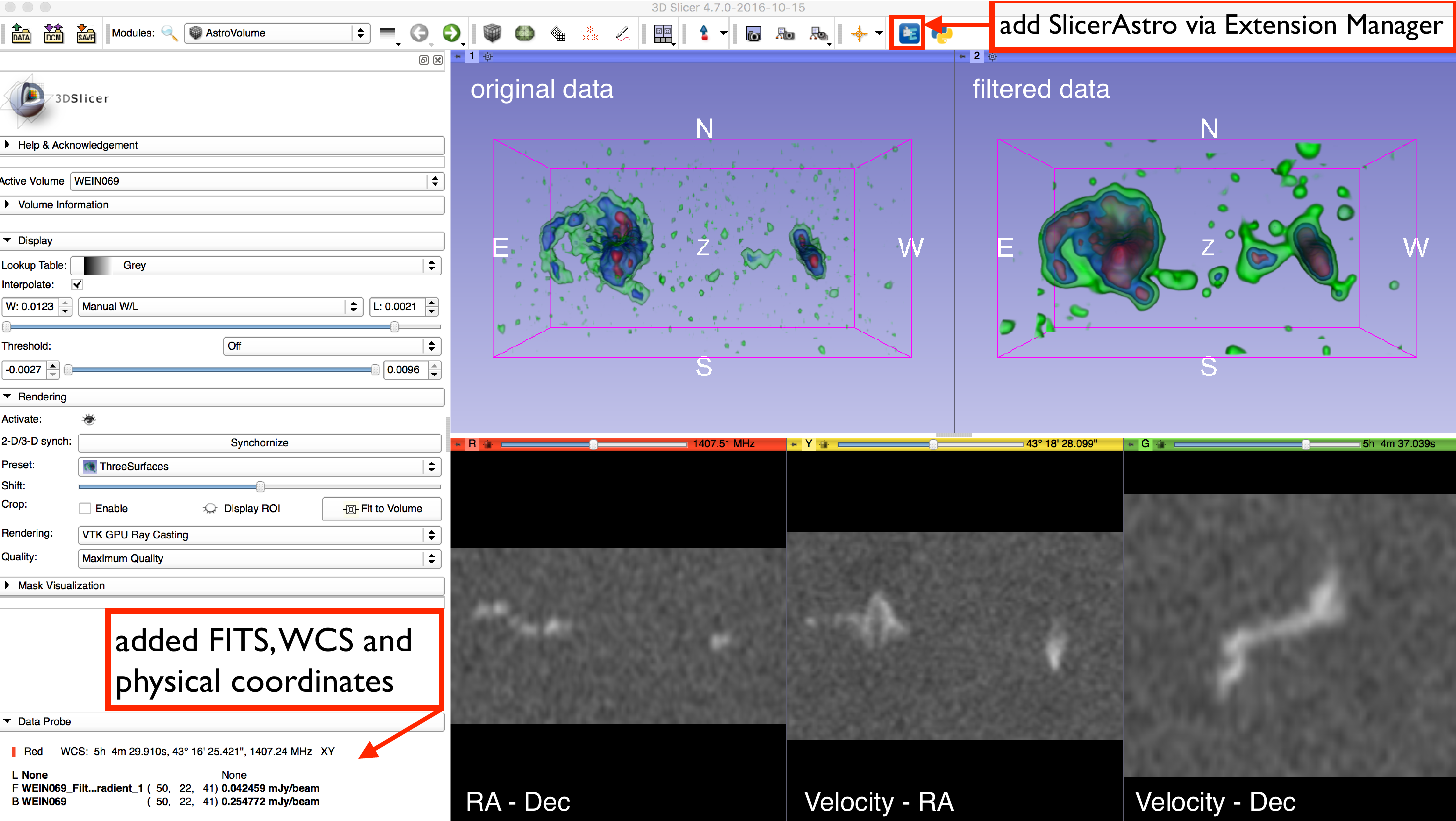} 
\vspace*{0.2 cm}
 \caption{Screenshot of a full view of $\tt{SlicerAstro}$ in $\tt{3DSlicer}$.
 The left panel contains widgets with various control functions described in the text, the right panel is a 5 frame layout showing linked 3D views of two data sets at the top and three linked 2D views of the left data set at the bottom. Additional labeling indicates the location of the extension manager control button (for adding packages) and the probe region (lower left) where the astronomical coordinates are displayed of the cursor position in the bottom 2D displays.}
   \label{fig1}
\end{center}
\vspace*{0.2 cm}
\end{figure}

\begin{figure}[tp]
\vspace*{0.2 cm}
\begin{center}
 \includegraphics[width=4.2in]{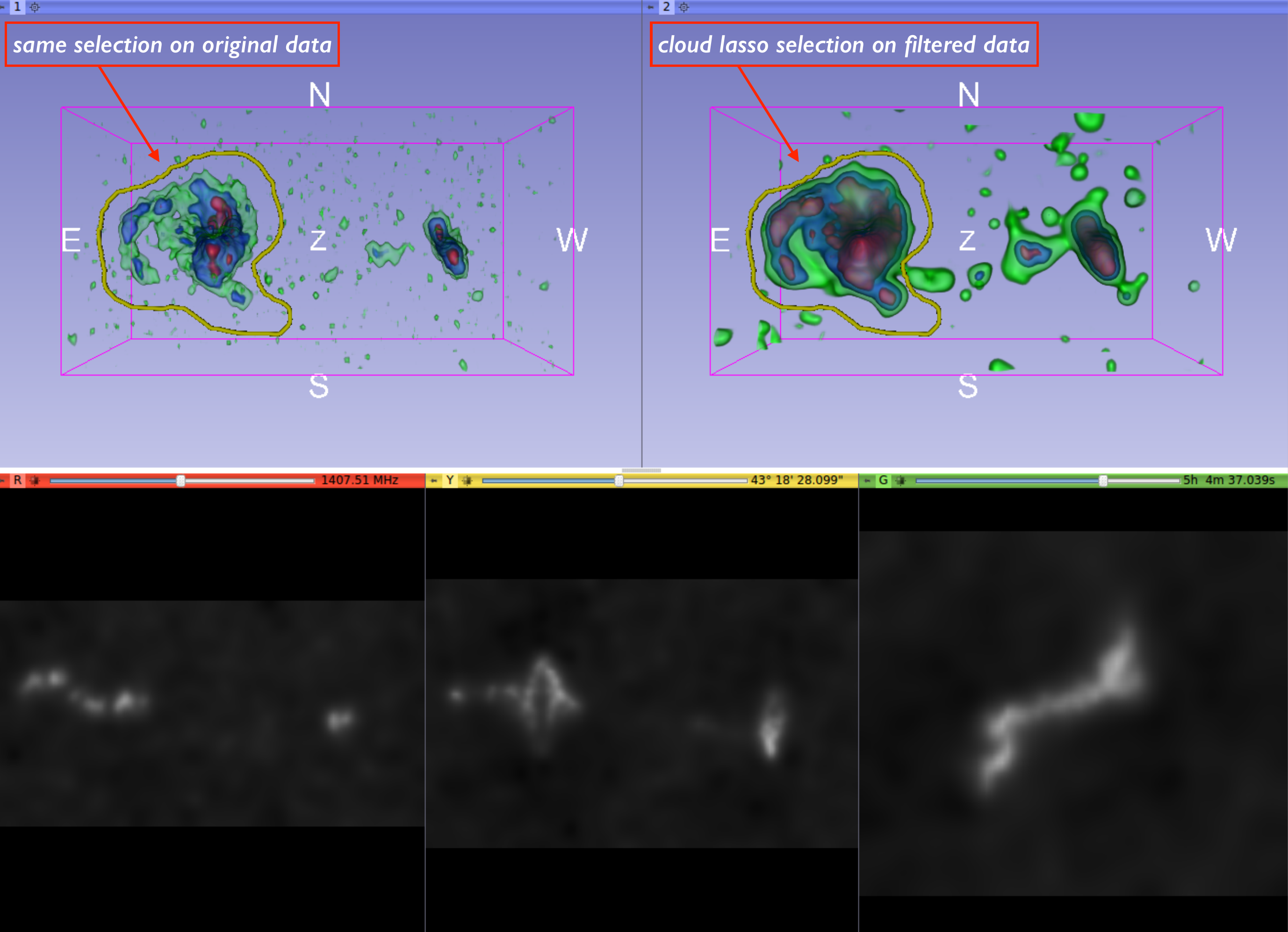} 
\vspace*{0.2 cm}
 \caption{Screenshot of the $\tt{CloudLasso}$ selection in $\tt{SlicerAstro}$. The selection as drawn by the user in the top right panel with the filtered data is indicated by the yellow line. The selection is simultaneously displayed in the top left panel displaying the original, unfiltered data. For the final selection see Figure \ref{fig3}}
   \label{fig2}
\end{center}
\vspace*{0.2 cm}

\vspace*{0.2 cm}
\begin{center}
 \includegraphics[width=4.2in]{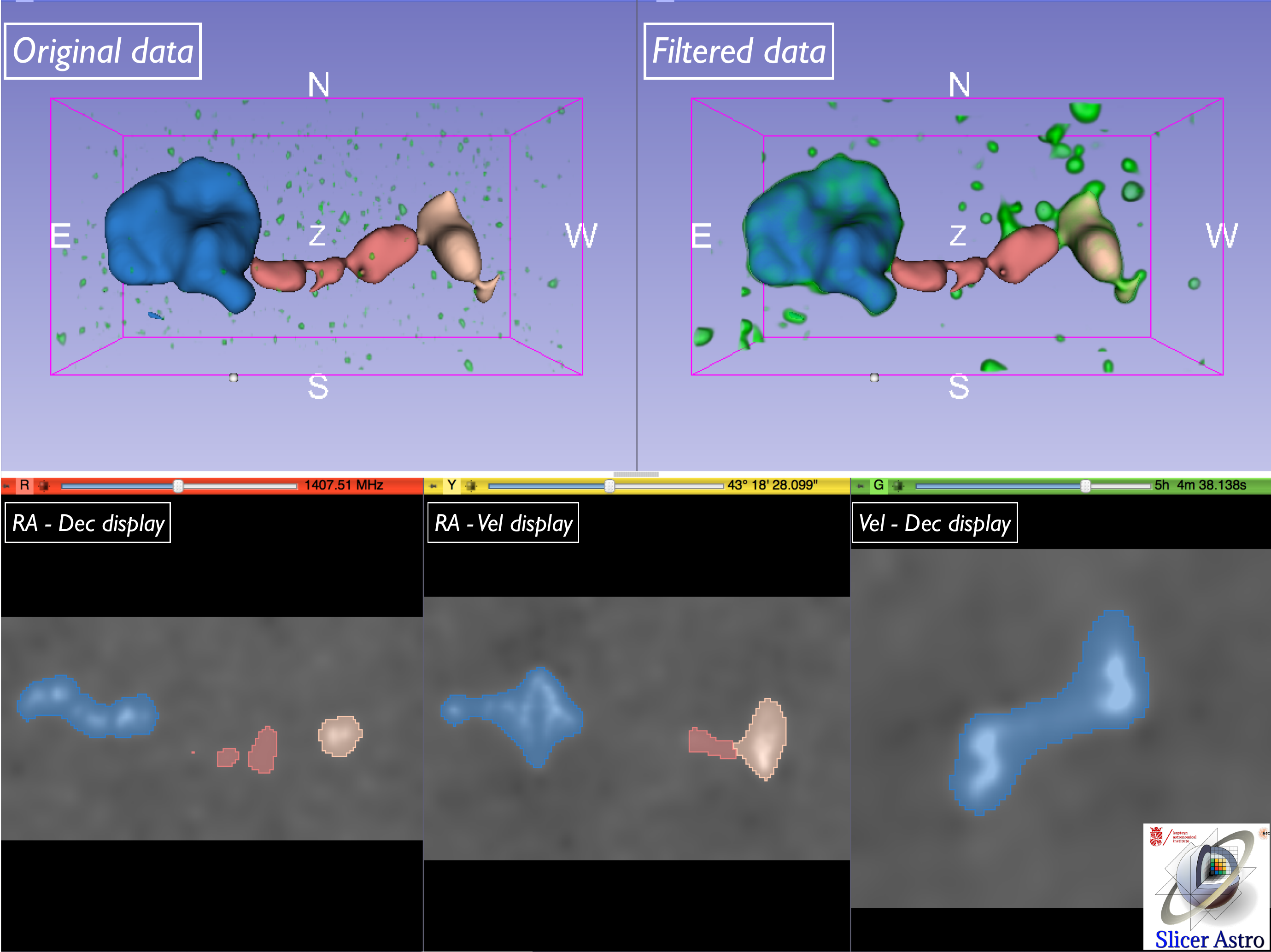} 
\vspace*{0.2 cm}
 \caption{Screenshot showing three selections, highlighted in blue, ochre and red respectively, with the $\tt{CloudLasso}$ selection tool. The blue segmentation shows the selection in the $\tt{CloudLasso}$ region indicated in Figure \ref{fig2}. The ochre and red segmentations are separate selections of the second galaxy and the  H\,{\small I} structure between the galaxies respectively. Both the two top 3-D displays and the bottom three 2-D displays show all three selections simultaneously. They can be switched on or off separately.}
   \label{fig3}
\end{center}
\end{figure}
\begin{enumerate}[I)]
\item visualization of astronomical data cubes using the FITS data format;
\item active use of proper astronomical world coordinates and physical units;
\item interactive (semi)automatic smoothing in all three dimensions;
\item interactive 3-D selection of H\,{\small I} sources;
\item interactive H\,{\small I} data modelling coupled to visualization.\newline
\end{enumerate}

Figure \ref{fig1} shows the basic layout of $\tt{SlicerAstro}$ in the $\tt{3DSlicer}$ environment. The left panel includes widgets 
for changing the colour transfer function for 2-D displays and for changing the properties of the volume rendering, In addition it includes interface widgets to control the display proprieties (e.g., user interaction to rotate the 3-D view) and a window for displaying the properties of a data probe coupled to the cursor in the linked 2-D views. 
The rest of the display in Figure  \ref{fig1} can be used to display the data in a number of user specified ways, ranging from a single 3D view to a combination of linked views as presented here.  In Figure  \ref{fig1} the arrangement of views consists of two linked 3D views of two different versions of the same data set (the original data to the left and the filtered data to the right) at the top and three linked 2D views at the  bottom displaying the original data in three slices: right ascension versus declination, velocity versus right ascension and velocity versus declination. At the top of these three 2D displays there is a slide bar which allows moving through the third coordinate, i.e. velocity, declination and right ascension respectively. 

\subsection{Filtering}

In order to find and enhance faint and extended features in H\,{\small I} data one usually lowers the resolution in right ascension, declination and/or velocity  to increase the signal-to-noise. The optimal shape of the smoothing kernel is of course the one that follows the 3D structure of the faint features in the data such that it is just resolved in each of the three dimensions. The exact shape is, however, not known a priory, so that in practice one often uses a set of different kernels and then inspects the results to make a final decision about which one suits best. 

\cite[Punzo et al. (2016)]{2016A&C....17..163P} have investigated this problem and showed that there is an optimum filtering algorithm that works very effectively on  H\,{\small I} data.  This filter is an adaptive filter, based on the intensity-gradient driven filter of \cite[Perona \& Malik (1990)]{perona&malik1990}, which appeared to be quite effective on H\,{\small I} data with well chosen default parameters. This filter, also known as {\it anisotropic diffusion}, does preserve the original resolution for regions where the signal-to-noise is high, but adapts to lower resolutions in regions where the signal-to-noise becomes very low but where there still is coherence in the data on larger spatial and/or velocity scales.  This method has been implemented in $\tt{SlicerAstro}$, in addition to the general Gaussian smoothing. The left 3D panel in Figure  \ref{fig1} shows a version of the high resolution data displayed at the left after application of this intensity-gradient driven filter.

\subsection{3-D data selection}

Having an optimized 3-D selection technique is still a problem that has not been solved satisfactorily, not only in astronomy, but also not in medical and other visualization applications. 
Moreover, the optimal selection technique
highly depends on the nature of the data, i.e. the complexity of the 3-D data structure. Requirements for a 3-D selection tool
are interactivity and minimal number of user-operations
for achieving the selection (i.e. user-friendliness).   
\cite[Yu et al. (2012a)]{Yu2012a} and \cite[Yu et al. (2012b)]{Yu2012b} provide a review of the state-of-the-art 3-D selection algorithms. In $\tt{SlicerAstro}$ we implemented the $\tt{CloudLasso}$ technique \cite{Yu2012a} for 3-D selection. 
The $\tt{CloudLasso}$ selection uses the
Marching Cubes (MC) algorithm (\cite[Wyvill et al. 1986]{Wyvill1986}, \cite[Lorensen \& Cline 1987]{Lorensen1987}) for the identification 
of regions of voxels with signal inside a user-drawn lasso, i. e. $\tt{CloudLasso}$ is
a lasso-constrained Marching Cubes method. The MC method allows
spatial selection of individual structures with a specified signal-to-noise ratio 
(usually $\gtrsim 3$) within a lasso region,
even if these lie visually one behind one another, without including the
noisy regions in between.  A detailed description of the $\tt{CloudLasso}$ selection can be found in Punzo et al. (2017, in preparation).

Figure \ref{fig2} shows a region selected with the $\tt{CloudLasso}$. The selection in Figure \ref{fig2} has been done on the filtered version of the data (top right panel) and is indicated by the closed yellow line. In this example the selection is automatically applied to the original data  shown  in the top left panel. The actual selection using the MC algorithm is shown in Figure \ref{fig3} as the region highlighted in blue. The selection (defined as a segmentation structure in $\tt{3DSlicer}$) is also shown in the three lower slices in Figure \ref{fig3}, highlighted in the same blue colour. In addition Figure \ref{fig3} shows two other selections made with the $\tt{CloudLasso}$: a red selection around the  faint H\,{\small I} bridge connecting the two galaxies in this example and a ochre selection highlighting the second galaxy. Also these two selections are shown in the three slices at the bottom. The $\tt{CloudLasso}$ implementation allows interactive modification of these selections .

The selections are saved in $\tt{3DSlicer}$ as segmentation structures, but can be exported as FITS files in the form of masks.  All voxels in such a FITS file are zeros, except for the ones within a segmentation structure. These are numbered 1, 2, 3 , etc. following the numbering of the segmentations. Conversely a FITS file with masks, produced by another programme, e.g. $\tt{SoFiA}$ (\cite[Serra et al. 2015]{2015MNRAS.448.1922S}), can be read into $\tt{SlicerAstro}$, be displayed and modified using the $\tt{CloudLasso}$ mechanism.

\subsection{Model aided selection}

Rather than selecting regions by hand, one can use a model representing the data to highlight the parts of the 3-D structure that fit the model well and separate those from the regions in the 3-D data volume that do not fit the data. A simple example is the following (see Figure 5 in \cite[Punzo et al. 2015]{2015A&C....12...86P}): one can use a classical tilted ring model to describe the regularly rotating H\,{\small I} disk of an observed galaxy and use this model to highlight all parts of the data that fit the model well. All data that does not fit the model, i.e. deviates from regular rotation, can then easily be viewed as it will not be highlighted, i.e. have a contrasting colour. The full 3-D rendering capabilities of $\tt{SlicerAstro}$ then allow the user to easily inspect the full 3-D structure of the H\,{\small I} that exhibits anomalous velocities.

This capability is being implemented in $\tt{SlicerAstro}$ as well, using the automated 3-D fitting routine $\tt{^{\rm3D}\,Barolo}$ \cite[(Di Teodoro \& Fraternali 2015)]{2015MNRAS.451.3021D} for fitting the regularly rotating H\,{\small I} disks in galaxies. This implementation provides both a (partly interactive) modelling capability and allows separating the H\,{\small I} with regular kinematics from the H\,{\small I} with anomalous kinematics in the data cubes of observed galaxies. The data used for the model fitting can be preselected using the $\tt{CloudLasso}$ procedure described above.

In this overview we have presented a new extension, $\tt{SlicerAstro}$, that has been added to the existing $\tt{3DSlicer}$ platform, developed originally for scientific visualisation of medical data.  $\tt{SlicerAstro}$ is not only capable of reading FITS format data, but also understand and handles astronomical world coordinates and physical units. The first few capabilities implemented in $\tt{SlicerAstro}$ and described here are filtering, 3-D selection and model aided exploration of H\,{\small I} data.  The system is still under development and we expect to add more features in the future. User feedback is greatly appreciated and can be directed to the first two authors of this paper or be  posted on https://github.com/Punzo/SlicerAstro.

\section{Acknowledgments}
D. Punzo and J.M van der Hulst gratefully acknowledge support from the European 
Research Council under the European Union's Seventh Framework Programme 
(FP/2007-2013)/ERC Grant Agreement nr. 291531.


\begin{thebibliography}{}


\bibitem[Calabretta \& Greisen(2002)]{2002A&A...395.1077C} Calabretta, M.~R., \& Greisen, E.~W.\ 2002, A\&A, 395, 1077 

\bibitem[Duffy et al.(2012)]{2012MNRAS.426.3385D} Duffy, A.~R., Meyer, M.~J., Staveley-Smith, L., et al.\ 2012, MNRAS, 426, 3385 


\bibitem[Di Teodoro \& Fraternali(2015)]{2015MNRAS.451.3021D} Di Teodoro, E.~M., \& Fraternali, F.\ 2015, MNRAS 451, 3021 

\bibitem[Gooch(1996)]{1996ASPC..101...80G} Gooch, R.\ 1996, Astronomical Data Analysis Software and Systems V, 101, 80 

\bibitem[Greisen \& Calabretta(2002)]{2002A&A...395.1061G} Greisen, E.~W., \& Calabretta, M.~R.\ 2002, A\&A, 395, 1061 

\bibitem[Greisen et al.(2006)]{2006A&A...446..747G} Greisen, E.~W., Calabretta, M.~R., Valdes, F.~G., \& Allen, S.~L.\ 2006, A\&A, 446, 747 


\bibitem[Hassan \& Fluke(2011)]{2011PASA...28..150H} Hassan, A., \& Fluke, C.~J.\ 2011, Publications of the Astronomical Society of Australia, 28, 150 

\bibitem[Johnston et al.(2008)]{2008ExA....22..151J} Johnston, S., Taylor, R., Bailes, M., et al.\ 2008, Experimental Astronomy, 22, 151 

\bibitem[Koribalski(2012)]{2012PASA...29..359K} Koribalski, B.~S.\ 2012, Publications of the Astronomical Society of Australia, 29, 359

\bibitem[Lorensen \& Cline (1987)]{Lorensen1987}
Lorensen, W.E. \& Cline, H.E. 1987 , SIGGRAPH Comput. Graph. vol. 21, nr. 4, 163

\bibitem[Norris(1994)]{1994ASPC...61...51N} Norris, R.~P.\ 1994, Astronomical Data Analysis Software and Systems III, 61, 51 

\bibitem[Oosterloo et al.(2010)]{2010iska.meetE..43O} Oosterloo, T., Verheijen, M., \& van Cappellen, W.\ 2010, ISKAF2010 Science Meeting, 43 

\bibitem[Perona \& Malik (1990)]{perona&malik1990} Perona, P \& Malik, J.\  
1990, IEEE, 12, 629
  
\bibitem[Punzo et al.(2016)]{2016A&C....17..163P} Punzo, D., van der Hulst, J.~M., \& Roerdink, J.~B.~T.~M.\ 2016, Astronomy and Computing, 17, 163 

\bibitem[Punzo et al.(2015)]{2015A&C....12...86P} Punzo, D., van der Hulst, J.~M., Roerdink, J.~B.~T.~M., et al.\ 2015, Astronomy and Computing, 12, 86 

\bibitem[Sancisi et al.(2008)]{2008A&ARv..15..189S} Sancisi, R., Fraternali, F., Oosterloo, T., \& van der Hulst, T.\ 2008, A\&A Review, 15, 189 

\bibitem[Serra et al.(2015)]{2015MNRAS.448.1922S} Serra, P., Westmeier, T., Giese, N., et al.\ 2015, MNRAS, 448, 1922 

\bibitem[Verheijen et al.(2009)]{2009pra..confE..10V} Verheijen, M., Oosterloo, T., Heald, G., \& van Cappellen, W.\ 2009, Panoramic Radio Astronomy: Wide-field 1-2 GHz Research on Galaxy Evolution, 10 

\bibitem[Wells et al.(1981)]{1981A&AS...44..363W} Wells, D.~C., Greisen, E.~W., \& Harten, R.~H.\ 1981, A\&A Suppl., 44, 363 

\bibitem[Whiting(2012)]{2012MNRAS.421.3242W} Whiting, M.~T.\ 2012, MNRAS, 421, 3242 

\bibitem[Wyvill et al. (1986)]{Wyvill1986} Wyvill, G., McPheeters, C. \& Wyvill, B. 1986, The Visual Computer, vol. 2, nr. 4, 227

\bibitem[Yu et al. (2012)]{Yu2012a} Yu, L., Efstatiou, K., Isenberg, P. \& Isenberg, T. 2012, IEEE Transactions on Visualization and Computer Graphics, vol. 18, nr. 12, 2245

\bibitem[Yu et al. (2012)]{Yu2012b}Yu, L., Efstatiou, K., Isenberg, P. \& Isenberg, T. 2012, IEEE Transactions on Visualization and Computer Graphics, vol. 22, nr. 1, 1077

\end{thebibliography}
\end{document}